# Microwave Quantum Radar using a Josephson Traveling Wave Parametric Amplifier

1st Patrizia Livreri
*Dept. of Engineering*
*University of Palermo, CNIT*
Palermo, Italy
patrizia.livreri@unipa.it

2nd Emanuele Enrico
*Quantum Metrology Division*
*INRIM*
Torino, Italy
e.enrico@inrim.it

3rd Luca Fasolo
*Dept. of Electronics and Telecom,*
*Polytechnique of Turin*
Torino, Italy
luca_fasolo@polito.it

4th Angelo Greco
*Quantum Metrology Division*
*INRIM*
Torino, Italy
a.greco@inrim.it

5th Alessio Rettaroli
*INFN*
Frascati, Italy
alessio.rettaroli@lnf.infn.it

6th David Vitali
*School of Science and Technology*
*Università di Camerino*
Camerino, Italy
david.vitali@unicam.it

7th Alfonso Farina
*AF Consulting*
Roma, Italy
alfonso.farina@outlook.it

8th Com. Francesco Marchetti
*Secretariat General of Defence and National Armaments Directorate*
*Ministero della Difesa*
Roma, Italy
sgd@sgd.difesa.it

9th A. Sq. Dario Giacomin
*Secretariat General of Defence and National Armaments Directorate*
*Ministero della Difesa*
Roma, Italy
csp.vicesgdna@sgd.difesa.it

*Abstract*— Detection of low-reflectivity objects can be improved by the so-called quantum illumination procedure. However, quantum detection probability exponentially decays with the source bandwidth. The Josephson Parametric Amplifiers (JPAs) technology utilized as a source, generating a pair of entangled signals called two-mode squeezed vacuum states, shows a very narrow bandwidth limiting the operation of the microwave quantum radar (MQR). In this paper, for the first time, a microwave quantum radar setup based on quantum illumination protocol and using a Josephson Traveling Wave Parametric Amplifier (JTWPA) is proposed. Measurement results of the developed JTWPA, pumped at 12 GHz, show an ultrawide bandwidth equal to 10 GHz at X-band making our MQR a promising candidate for the detection of stealth objects.

*Keywords—Microwave quantum radar, Josephson traveling parametric amplifier, quantum microwave illumination, two-mode squeezing, non-classical light source, quantum signals, entanglement.*

## I. Introduction

The concept of quantum illumination, initially proposed by S. Lloyd [1] and then improved and extended via Gaussian states [2], is a subject of intense research activity, representing one of the very few examples in which "quantum advantage" is achieved in noise-dominated experimental contexts. Suitable strategies have to be introduced to protect entanglement from the detrimental effects of the environment. Quantum lighting, even if based on entangled radiation sources, is capable to exponentially reduce the probability of error in the detection of a target compared to the case of traditional sources, precisely in the limit of low signal to noise ratio (SNR $\leq 0.01$) where the input signal is no longer entangled. Quantum illumination schematizes the problem in a binary way, i.e., the discrimination between the presence or absence of a target, and considers the limiting case where the signal at the receiver is dominated by noise, meaning that the probability of receiving radiation reflected from the target is low. In general, both the radiation source and the measurement on the receiver need to be optimized and traditional measurements is limited. In the quantum case, the ideal source generates entangled states such as two-mode squeezed vacuum states [3, 4], which can be produced by adjusting the parameters of the amplifiers (either optical or exploiting the Josephson effect for microwaves), and the ideal receiver is the so-called FF-SFG [5, 6]. The quantum strategy provides an exponential reduction of the error probability $P_e$ in discrimination, namely a ~ 6 dB decrease in the error probability exponent compared to the optimal classical benchmark. Preliminary experimental demonstrations have recently been carried out, using entangled microwave sources based on Josephson parametric amplifiers revealed a target about 1 meter away in the laboratory, outside the cryogenic environment where the source was located [7, 8]. Although the detection is far from optimal, the experiments showed an improvement in SNR compared to a classic two-beam source (signal + idler) of chaotic type (radar noise). A further increase in SNR of about 1 dB (in the range of about 0.1 photons on average) compared to the classic optimal case was then achieved by a similar entangled source [9] and a digital phase conjugator receiver, avoiding the need for a quantum memory [10,11]. This experimental proof of principle indicates that alternative detection schemes are possible and that the ultimate gain limit of 6 dB in the error probability exponent can be reached in different ways. A new concept of one-dimensional metamaterial, with multiple Josephson junctions incorporated, promoting strong photon-photon-on-chip interactions has been recently demonstrated [12], allowing experimenters to design dispersion relationships that convey mixing interactions along d-guides [13,14]. These basic concepts and advanced technologies allow for the control and tuning of the wave mixing process. For example, a weak signal within a metamaterial can interact with a strong pump tone at a different frequency, activating the so-called parametric amplification [15]. The class of devices where these phenomena are promoted and engineered is commonly known as Josephson Traveling Wave Parametric Amplifiers (JTWPA). It has been shown that JTWPAs can act as quantum parametric amplifiers reaching the so-called quantum limit.

The ability to overcome the quantum limit is related to the so-called phase-sensitive amplification process, where the metamaterial can operate in degenerate mode (degenerate parametric amplifier, DPA), acting on two waves (signal and idler) at the same frequency amplifying and de-amplifying their quadratures, corresponding to quantum position and momentum. A broadband squeezed light source such as a TWJPA also supplies great advantage towards scalability, because operations can be parallelized by many pairs of strongly separated two-way compressed frequencies exploiting a single device.

In this work, starting from the operating scheme based on the "quantum illumination" protocol, an experimental set-up in view of microwave quantum radar realization using the superconducting parametric amplifier with Josephson traveling-wave junctions operating in an ultra-cryogenic environment, is proposed. The JTWPA design, development, and measurement results are here reported and discussed.

## II. QUANTUM ILLUMINATION

Quantum illumination is a photonic quantum remote sensing protocol in which shared entanglement between a signal probing a target region and a photonic mode (called "idler") stored locally in the emitting station is used to detect the presence of a weakly reflective lens with higher efficiency than a strategy that uses a single signal with the same energy, i.e. with the same average number of photons.

By preparing the idler modes in a two-mode squeezed vacuum state (TMSV), with a quantum illumination protocol it is possible to obtain a 6 dB improvement in the exponent of the probability error compared to a classic ladar type protocol (laser detection and ranging, i.e. detector and locator via laser), with the same energy. Surprisingly, and unlike other quantum sensor protocols for which the quantum advantage quickly disappears in the presence of noise and decoherence sources, this remarkable 6 dB theoretical advantage occurs just when the target is immersed in a background of thermal radiation of luminosity (ie energy per mode) $N\_B >> 1$, whereby the initial entanglement between signal and idle rapidly degrades until it is completely destroyed. This is precisely the most interesting feature of the quantum illumination protocol, namely that its quantum advantage over any other possible classical strategy is obtained precisely in situations of maximum noise.

The more general quantum illumination protocol foresees the preparation of the signal mode and the idler mode in an initial entangled state, which therefore will not be a product state of the idler I and the signal S:

$$|\psi\rangle_{IS} = \sum_{n} \sqrt{p_n} |\chi_n\rangle_I |n\rangle_S \qquad (1)$$

where, $|\chi\_n>\_I$ are the base states of the idler, and $|n>\_S$ are the base numbers (Fock states) of the signal.

The quantum correlation associated with the entanglement of this pure state is ensured by the linear superposition of the states of the bipartite system composed of signal and idler. In an illumination protocol, the receiving station performs a joint measurement on the reflected mode and the idler mode that minimizes the average probability of error P_e. It consists of the convex combination of the probability P_F of incorrect identification of the target (false positive) and the probability P_M of non-identification of the target (miss probability):

$$P\_e = \lambda P\_F + (1- \lambda)\ P\_M \qquad (2)$$

In particular, non classical ladar with a signal energy Ns can be associated with a probability of error lower than the value:

$$P_e^{cl} \simeq \exp\left[-\eta \mathcal{N}_S \left(\sqrt{N_B + 1} - \sqrt{N_B}\right)^2\right] \qquad (3)$$

where, η is the reflectance of the signal transmitted on the path of the return beam when the target is present.

On the contrary, the probability of error of the quantum illumination protocol using an initial entangled state idler-signal of the TMSV type is shown to be reduced well below the classical minimum value, up to the limit:

$$P_e^{TMSV} \simeq \exp(-\eta \mathcal{N}_S / N_B) \qquad (4)$$

In the presence of many background photons, a quantum illumination protocol allows a significant reduction in the probability of error both of non-identification and of incorrect identification of the target. In Fig. 1, obtained from [1], the basic protocol is schematically represented, in which the transmitting station prepares the entangled idler-signal state, sends the signal to the target region, holds the idler mode, and then carries out a measurement joint of the reflected signal and the idler mode. The various electromagnetic modes $a\hat{}\_S\hat{}((m))$, $a\hat{}\_I\hat{}((m))$, $a\hat{}\_B\hat{}((m))$, and $a\hat{}\_R\hat{}((m))$ indicate, respectively, the m-th quantum mode of the transmitted signal, of the idler, of thermal background, and of the reflected signal, while ϕ denotes the phase shift of the reflected mode with respect to the transmitted mode.

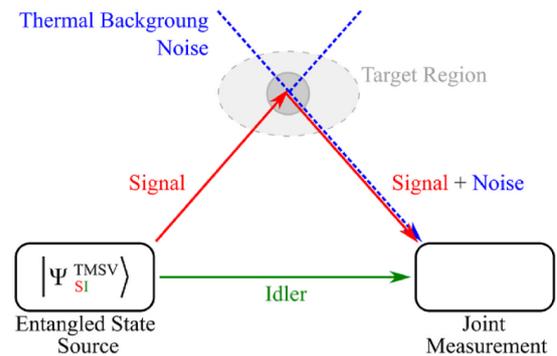

Fig. 1. Quantum illumination block schematic [1].

## III. MICROWAVE QUANTUM RADAR

A. *Schematic representation of microwave quantum radar set-up based on a JTWPA.*

A simplified block diagram of the radiation source for the proposed microwave Quantum Radar Two mode-squeezing (mQRTMS) set-up is shown in Fig. 2a. The main difference between our quantum radar setup and the classical microwave quantum radar is in the signal generation step [16,17]. The Josephson parametric amplifier (JPA) exploited in [9, 10, 11], the source of the TMSV entangled signal, is replaced by a JTWPA (shown in Fig. 2b) [18]. JTWPA is the core of the innovation in our system being the only component not commercially available at the present moment. RF-SQUIDs based JTWPA can be biased by a DC current in order to promote or suppress three-wave mixing (3WM) interactions by tuning the Josephson nonlinearity. In the 3WM working point, a quadratic dependence of the gain on the pump power is reached, so that quantum correlations of the output field are maximized with respect to the four-wave mixing (4WM) regime where linear gain-pump power relation occurs.

It has been shown that quantum correlations fingerprints survive when JTWPAs are located in a realistic environment characterized by non-zero thermal noise only when the amplifier itself expresses a high gain figure of merit [19].

In the next section, the cryogenic experimental setup and the JTWPA will be examined in detail.

The core of the experiment runs into a dry dilution refrigerator, manufactured by Leyden Cryogenics capable of reaching a base temperature of 15 mK and contains several stages with different cooling powers. The JTWPA is positioned at the bottom of the cryostat, in the coldest stage, in order to maintain the thermal noise contribution as low as possible. This level of cooling is not only necessary because of the superconducting nature of the JTWPA device, but it guarantees state-of-the-art thermal noise background radiation that otherwise would destroy the entanglement generated by the parametric amplification. A low-noise amplifier based on HEMT is introduced as the first stage of the chain of amplifiers, this latter having a negligible contribution in terms of noise of the readout setup according to the well known Friis Formula for a multistage amplifier.

The number of independent pulses $M$ emitted by a light source depends on its bandwidth $B$ and the total integration time $T$ via the simple relation $M = TB$. Quantum detection probabilities exponentially decays with the source bandwidth.

*B. Transmit Power*

The entire experimental setup is designed in order to guarantee optimal working conditions in the range 4-8 GHz, essentially limited by the bandwidth of the commercially available cryogenic circulators introduced to prevent the back-action of the HEMT on the JTWPA. It's worth mentioning that, in an ideal condition, the number of independent pulses M emitted by a light source depends on its bandwidth B and the total integration time $T$ via the simple relation $M=TB$. Quantum detection probabilities depend exponentially by the bandwidth of the light-source. The transmit power is approximately P=E*B, assuming one photon per mode, where E=hv is the single photon energy with h the Planck's constant and ν the photon frequency. Hence, the wider bandwidth gained with the JTWPA and the hosting cryogenic setup allows a transmit power value for a single quantum device higher than the value obtained with the JPA, limited by a bandwidth in the order of tens of MHz. That is, for a MQR using a JTWPA, the maximum transmit power for a single quantum transmit device is expected to be three orders of magnitude above the value of the MQR using JPA.

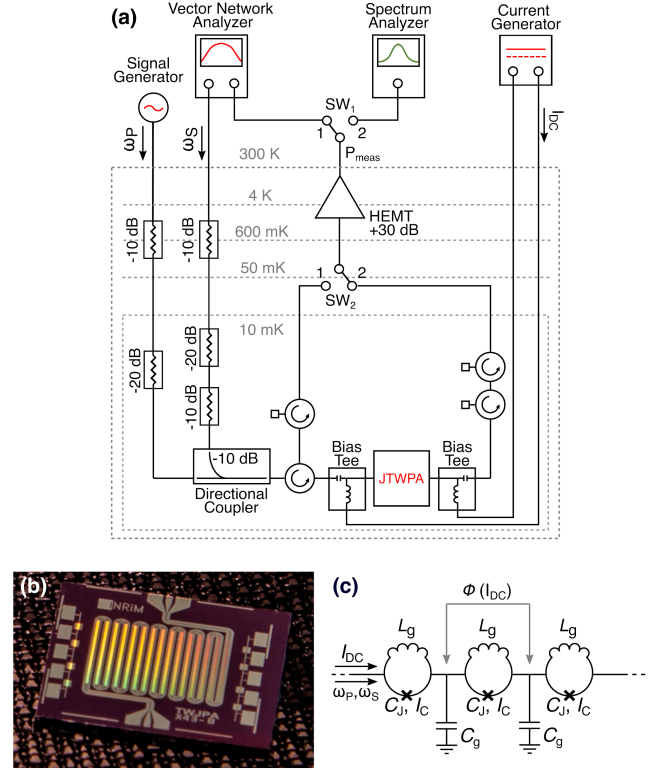

Fig. 2. (a) Schematic of the experimental setup used for the power and S-parameters measurements. The signal and pump tones, provided by a signal generator and a VNA, pass through several attenuation stages arriving at a directional coupler that inject the tones into the JTWPA after a first isolation stage provided by a circulator. An electromechanical switch (SW2) in the output line allows to perform reflection (position 1) or transmission (position 2) measurements. Another switch (SW1) allows to change the detector at room temperature between the VNA (position 1) and the spectrum analyzer (position 2). The output signal is amplified by a 30 dB HEMT placed on the 4K stage. A current generator supplies a DC current that is injected into the central conductor of the JTWPA through two bias tees. (b) Photograph of the JTWPA under test. (c) Circuit schematic of the Josephson metamaterial. The central conductor of the coplanar waveguide is made by a chain of rf-SQUIDs composed by a geometrical inductor Lg and a Josephson junction of critical current $I_c$ and capacitance $C_J$. The line is capacitively shunted to ground by ground capacitors $C_g$.

IV. JOSEPHSON TRAVELING WAVE PARAMETRIC AMPLIFIER

The Josephson Traveling Wave Parametric Amplifier (JTWPA) is a nanostructured superconducting device composed of the repetition of several Josephson junctions (JJs) embedded in a coplanar waveguide (CPW) as shown in Fig. 2c. The JJs constitute a uniqueness in the solid state physics, being the only known non dissipative nonlinear passive elements. In the context of MQI, these devices represent a source of non-classical light played by the nonlinear crystals in the QI. Microwave tones propagate through these artificial structures and exchange energy with other tones thanks to the intermodulation energy-preserving phenomena, generating entangled quantum states. It has been experimentally demonstrated how for a JTWPA the gain G can reach values of about 20 dB over an unprecedently demonstrated high bandwidth [8].

The Josephson metamaterial (shown in Figure 2b) is composed of 15 sections of coplanar waveguide embedding rf-SQUIDs connected by bended sections of CPW. The whole device contains 990 non-hysteretic rf-SQUIDs, characterised by values for the circuit parameters: ground capacitance $C_g$= 13.0 fF, geometrical inductance $L_g$= 45 pH, Josephson capacitance $C_J$= 25.8 fF and Josephson critical current $I_c$= 1.5 μA. The values of $L_g$ and $C_J$ have been obtained by means of finite elements simulations [20], while $I_c$ has been estimated taking into account the junction area (≈ 0.4μm, measured by means of Scanning Electron Microscopy) and the junction critical current density, estimated using an empirical curve, obtained after the switching current measurement of Josephson junctions fabricated with the same technique, which relates the junction critical current density with the oxidation time and oxygen pressure used for the creation of the oxide barriers.

## V. MEASUREMENT SET-UP AND RESULTS

The characterization of the realized rf-SQUID-based JTWPA in terms of its 3WM capabilities and gain evaluated via pump-on pump-off technique has been carried out.

Fig. 2a shows a scheme of the experimental setup used for the cryogenic characterization. In order to quantify nonlinear mixing effects, a 2 tones measurement is performed by supplying in input a weak signal tone (supplied by port 1 of an Agilent E5071C 300 kHz-20 GHz VNA) and a driving pump tone (coming from a Rohde&Schwarz SMA100B 8 kHz-20 GHz signal generator). Microwave signals enter to a dilution refrigerator and pass through several attenuation stages, getting to the metamaterial after passing a directional coupler (Mini-Circuits ZUDC10-02183-S+) and a first isolation stage provided by a circulator (LNF-CIC412A). The microwave tones are then detected at room temperature after passing through a High Electron Mobility Transistor (HEMT, (LNF-LNC620C)) amplifier placed on the 4 K stage, which provides 30 dB of amplification. The room temperature switch SW1 allows choosing, as a receiver, a spectrum analyzer (Signal Hound SM200B 100kHz-20 GHz) or port 2 of the VNA. This permits to perform power-spectra or scattering parameters measurements, respectively. The cryogenic electromechanical switch SW2 allows to adapt the setup for transmission or reflection measurements. In both configurations, the output microwave passes through an isolation stage realized by means of two circulators (LNF-CIC412A). A current generator (Keithley 6221) connected to the device via a couple of bias tees (Marki BT-0018) provides the DC current bias to the device.

Nonlinear effects generate idler tones that have different frequencies depending on the order of nonlinearity that causes them. A first characterization of the JTWPA has been performed measuring the power of the output idler tone $P_{Idler}$, generated via 3WM, as a function of the DC bias current $I_{DC}$ (Fig. 3), providing a $\nu_P$= 6.75 GHz driving pump tone, with three different power values (-90, -85, and -80 dBm), and a $\nu_S$= 3.3 GHz signal tone with $P_S$= −64 dBm power. The powers are here considered at the device input. For this mixing process the idler is generated at $\nu_I = \nu_P - \nu_S$= 3.5 GHz

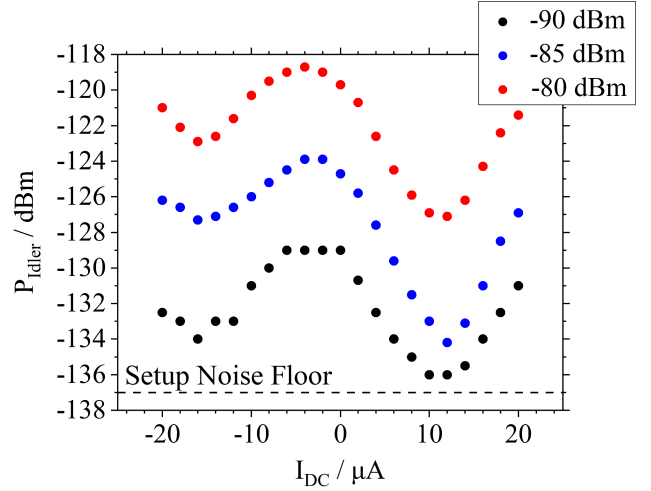

Fig. 3. Modulation of the output idler tone power ($P_{Idler}$) generated via 3WM as a function of the bias current $I_{DC}$. Here the JTWPA is excited with a signal tone at frequency $\nu_S$ = 3.3 GHz , for three different values of the driving pump tone at frequency $\nu_P$ = 6.75 GHz.

Due to the natural Kerr-nonlinearity [18, 21] of a rf-SQUID with no current bias one would expect that the 3WM idler would present a minimum at zero $I_{DC}$. Nonetheless Figure 2 (a) reports a shift of the minima that can be attributed to magnetic field trapping during the cooling phase of the dilution refrigerator. Moreover, the suppression of the 3WM idler tone via $I_{DC}$ is not complete since data reported in Figure 2(a) doesn't reach the noise floor of the setup represented by the dashed horizontal line for every $I_{DC}$ value. It has to be noticed that the modulation of the 3WM process here reported is limited to around 10 dB, since it is reasonably affected by nonidealities of the JTWPA and the surrounding environment.

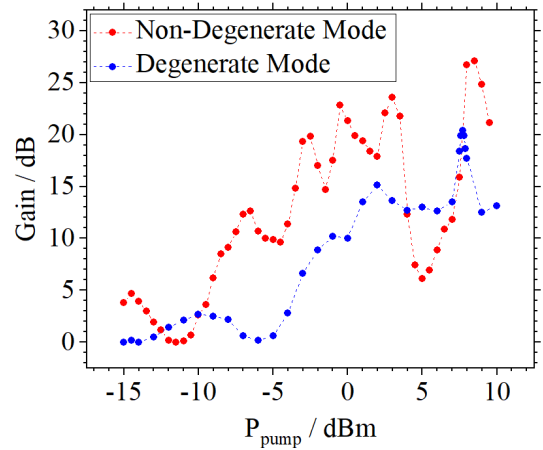

Fig. 4 Signal power Gain as a function of pump power ($P_{pump}$) for degenerate ($\nu_P$ = 18 GHz) and non-degenerate ($\nu_P$ = 13.4 GHz) regimes. Here the signal tone is kept at a fixed frequency $\nu_S$ = 9 GHz and the Gain is evaluated via pump-on-pump-off technique.

Even in this preliminary configuration, Figure 4 reports the crucial figure of merit of the JTWPA under study, being the gain, evaluated by means of pump-on-pump-off technique. There, the effect of the pump power (Ppump) on the signal gain for the two relevant regimes (degenerate and non-degenerate) of the amplifier is reported. The power gain induced by Josephson nonlinearities reaches values around 25 dB, showing that even in a broad range of interacting modes, the JTWPA acts as a metamaterial efficiently promoting 3WM processes, the key enabler for the preparation of two-mode squeezed vacuum states [18].

## VI. CONCLUSIONS

In this paper, a novel Microwave Quantum Radar (MQR) setup based on Quantum Illumination Protocoll using a Josephson Traveling Wave Parametric Amplifier (JTWPA) is proposed. Measurement results of the designed and developed JTWPA pumped at ωp=12 GHz show an ultrawide signal bandwidth B = 10 GHz. As quantum detection probability exponentially decays with the source bandwidth, the proposed ultra-wideband MQR setup represents a good candidate to detect stealth objects.

## ACKNOWLEDGMENT

The scientific and technological research presented in this work is co-financed by the Italian Ministry of Defence within the framework of the Master Agreement between the Secretariat General of Defence and National Armaments Directorate (SGD/NAD) and the National Inter-University Consortium for Telecommunications (CNIT) within the Quantum Radar Project Agreement 01/2021 (CIG: ZAE31B3A7C).
This research has been also supported by DARTWARS, a project funded by Istituto Nazionale di Fisica Nucleare (INFN, National Scientific Committee 5), by the SUPERGALAX project in the framework of the European Union (EU) Horizon 2020 research and innovation programme (H2020, FETOPEN-2018-2020 call), and by the Joint Research Project PARAWAVE of the European Metrology Programme for Innovation and Research (EMPIR). This project (PARAWAVE) received funding from the EMPIR programme co-financed by the Participating States and from the European Union's Horizon 2020 research and innovation programme.

## REFERENCES

[1] S. Lloyd, Enhanced sensitivity of photodetection via quantum illumination. Science 321, 1463–1465 (2008).

[2] S.-H. Tan, B. I. Erkmen, V. Giovannetti, S. Guha, S. Lloyd, L. Maccone, S. Pirandola, J. H. Shapiro, Quantum illumination with Gaussian States. Phys. Rev. Lett. 101, 253601 (2008).

[3] G. De Palma, J. Borregaard, The minimum error probability of quantum illumination, Phys. Rev. A 98, 012101 (2018).

[4] Q. Zhuang, Z. Zhang, J. H. Shapiro, Optimum mixed-state discrimination for noisy entanglement-enhanced sensing. Phys. Rev. Lett. 118, 040801 (2017).

[5] Piccolini, M., Nosrati, F., Compagno, G., Livreri, P., Morandotti, R. & Lo Franco, R. Entanglement robustness via spatial deformation of identical particle wave functions, Entropy **23**, 708 (2021).

[6] Q. Zhuang, Z. Zhang, J. H. Shapiro, Entanglement-enhanced Neyman–Pearson target detection using quantum illumination, JOSA B 34, 1567-1572 (2017).

[7] Sweeny M, Mahler R. A travelling-wave parametric amplifier utilizing Josephson junctions. IEEE Transactions on Magnetics. 1985;21,2:654-655. DOI:10.1109/TMAG.1985.1063777

[8] Macklin C, OBrien K, Hover D, Schwartz ME, Bolkhovsky V, Zhang X, Oliver, WD, Siddiqi I. A nearquantum-limited Josephson traveling-wave parametric amplifier. Science. 2015;350,6258:307-310. DOI: 10.1126/science.aaa8525

[9] S. Guha, B. I. Erkmen, Gaussian-state quantum-illumination receivers for target detection.Phys. Rev. A 80, 052310 (2009).

[10] S. Barzanjeh, S. Pirandola, D. Vitali, J. M. Fink, Microwave quantum illumination using a digital receiver. Sci. Adv. 6, eabb0451 (2020).

[11] S. Barzanjeh, S. Pirandola, D. Vitali, J. Fink, Microwave quantum illumination with a digital phase-conjugated receiver, IEEE National Radar Conference - Proceedings, 2020, 2020-September, 9266397.

[12] Dell'Anno, F., De Siena, S., & Illuminati, F., Multiphoton quantum optics and quantum state engineering. Physics Reports 428, 53-168 (2006).

[13] Palma, G. M. & Knight, P. L. Phase-sensitive population decay: the two-atom dicke model in a broadband squeezed vacuum. Phys. Rev. A 39, 1962–1969 (1989).

[14] Gómez, A. V., Rodriguez, F. J., Quiroga, L. &Garca-Ripoll, J. J. Entangled microwaves as a resource for entangling spatially separate solid-state qubits: superconducting qubits, nvcenters and magnetic molecules. Preprint at arXiv:1512.00269 (2015).

[15] Adesso, G., & Illuminati, F. Equivalence between entanglement and the optimal fidelity of continuous variable teleportation. Phys. Rev. Lett. 95, 150503 (2005).

[16] Bhashyam Balaji, Marco Frasca and Alfonso Farina, "Quantum Radar Research: A Snapshot in Time", IEEE Systems Magazine, AES, Special Issue on QR, April 2020. Pp. 74-76.

[17] D. Luong, C. W. S. Chang, A. M. Vadiraj, A. Damini, C. M. Wilson and B. Balaji, "Receiver Operating Characteristics for a Prototype Quantum Two-Mode Squeezing Radar," in *IEEE Transactions on Aerospace and Electronic Systems*, vol. 56, no. 3, pp. 2041-2060, June 2020, doi: 10.1109/TAES.2019.2951213.

[18] L Fasolo, A Greco, E Enrico, F Illuminati, R Lo Franco, D Vitali, P Livreri, "Josephson Traveling Wave Parametric Amplifiers as non-classical light source for Microwave Quantum Illumination", Measurements Sensors, vol.18, DOI: 10.1016/j.measen.2021.100349

[19] L Fasolo, et al. "Bimodal Approach for Noise Figures of MeritEvaluation in Quantum-Limited JosephsonTraveling Wave Parametric Amplifiers" arXiv, 30 sept. 2021.

[20] A. L. Grismo, A. Blais, "Squeezing and quantum state engineering with Josephson travelling wave amplifiers", npj Quantum Inf 3, 20 (2017). https://doi.org/10.1038/s41534-017-0020-8.

[21] L. Fasolo et al., "Bimodal Approach for Noise Figures of Merit Evaluation in Quantum-Limited Josephson Traveling Wave Parametric Amplifiers", arXiv preprint arXiv:2109.14924.

[22] M.Renger et al .,"Beyond the standard quantum limit of parametric amplification". Preprint arXiv:2011.00914 v3 [quant-ph].